\documentclass[lettersize,journal]{IEEEtran}
\usepackage{amsmath,amsfonts}
\usepackage{algorithmic}
\usepackage{algorithm}
\usepackage{array}
\usepackage[caption=false,font=normalsize,labelfont=sf,textfont=sf]{subfig}
\usepackage{textcomp}
\usepackage{stfloats}
\usepackage{url}
\usepackage{verbatim}
\usepackage{graphicx}
\usepackage{cite}
\usepackage{hyperref}
\usepackage{booktabs}
\usepackage{colortbl}
\usepackage[html]{xcolor}
\usepackage{svg}
\usepackage{amsmath}

\def \x {\mathbf{x}}
\def \y {\mathbf{y}}
\def \z {\mathbf{z}}

\def \I {\mathbf{I}}
\def \W {\mathbf{W}}

\newcommand{\ul}[5]{{\underline{#1}}{\underline{#2}}{\underline{#3}}{\underline{#4}}{\underline{#5}}}

\begin{document}

\title{Domain Specific Denoising Diffusion Probabilistic Models for Brain Dynamics}

\author{Yiqun Duan,~\IEEEmembership{Member,~IEEE},  Jinzhao Zhou, Zhen Wang, Yu-Cheng Chang~\IEEEmembership{Member,~IEEE}, Yu-Kai Wang~\IEEEmembership{Senior Member,~IEEE}, Chin-Teng Lin~\IEEEmembership{Fellow,~IEEE}
\thanks{Yiqun Duan, Jinzhao Zhou, Yu-Cheng Chang, Yu-Kai Wang and Chin-Teng Lin are with Human-centric Artificial Intelligence Centre, Australia Artificial Intelligence Institute, School of Computer Science, the University of Technology Sydney, Zhen Wang is with the School of Computer Science, University of Sydney}
\thanks{Manuscript submitted May 15, 2023;}}

\markboth{Journal of \LaTeX\ Class Files,~Vol.~14, No.~8, August~2021}%
{Shell \MakeLowercase{\textit{et al.}}: A Sample Article Using IEEEtran.cls for IEEE Journals}


\maketitle

\begin{abstract}

The differences in brain dynamics across human subjects, commonly referred to as human artifacts, have long been a challenge in the field, severely limiting the generalizability of brain dynamics recognition models. Traditional methods for human artifact removal typically employ spectrum filtering or blind source separation, based on simple prior distribution assumptions, which ultimately constrain the capacity to model each subject's domain variance.
In this paper, we propose a novel approach to model human artifact removal as a generative denoising process, capable of simultaneously generating and learning subject-specific domain variance and invariant brain signals. We introduce the Domain Specific Denoising Diffusion Probabilistic Model (DS-DDPM), which decomposes the denoising process into subject domain variance and invariant content at each step. By incorporating subtle constraints and probabilistic design, we formulate domain variance and invariant content into orthogonal spaces and further supervise the domain variance with a subject classifier.
This method is the first to explicitly separate human subject-specific variance through generative denoising processes, outperforming previous methods in two aspects: 1) DS-DDPM can learn more accurate subject-specific domain variance through domain generative learning compared to traditional filtering methods, and 2) DS-DDPM is the first approach capable of explicitly generating subject noise distribution. Comprehensive experimental results indicate that DS-DDPM effectively alleviates domain distribution bias for cross-domain brain dynamics signal recognition.
\end{abstract}


\section{Introduction}
\IEEEPARstart{T}{he} recognition of human brain dynamics signals such as electroencephalogram (EEG)~\cite{nunez2006electric} and Event-related Potential (ERP)~\cite{picton1995recording} is of vital importance for the non-invasive brain-computer interface. 
However, the brain signal distribution from different human subjects exhibits severe distributional differences~\cite{henry2006electroencephalography,jiang2019removal}, which means the recognition model~\cite{lawhern2018eegnet,AAAI22_CL,duan2019learning,duan2022cross} trained for a set of human subjects might not be efficient for other unknown human subjects. 
This weakens the generalized ability of deep learning-based models. 

Previous efforts for human artifacts removal~\cite{jiang2019removal} have sought to alleviate the aforementioned problem and can be categorized into four main approaches: 1) Regression~\cite{al2018complexity}, 2) Blind Source Separation (BSS)\cite{sweeney2012artifact,somers2016removal}, 3) Empirical-mode Decomposition (EMD), and 4) Wavelet Transform algorithm combined with their hybrid methods\cite{james2004independent}. Among these methods, Regression, and BSS are the most commonly used.
Classic regression methods~\cite{sweeney2012artifact} operate under the assumption that each channel is a sum of clean EEG data and a proportion of artifacts given by known reference signals~\cite{hillyard1970eye,wallstrom2002correction} through a set of regressed transmission factors. However, these methods only learn transmission factors instead of directly learning domain variance.
BSS methods, such as Independent Component Analysis (ICA), decompose observed signals into independent components (ICs)~\cite{somers2016removal} from linear mixtures of cerebral and artifactual sources. Current ICA methods still require reference signal information for artifact separation and do not consider the domain variance between different human subjects.

Previous human artifact removal methods generally assume domain variance as simple prior distributions, limiting their capacity to model domain variance for each subject. We propose to model human artifact removal as a domain-specific denoising process and perform denoising by generating subject-specific domain variance and invariant brain signals through the denoising process. Our intuition is to introduce learned subject-specific domain variance instead of a simple prior distribution.
To achieve this, we propose modeling human artifact removal as a domain-specific denoising process and performing denoising by generating subject-specific domain variance and invariant brain signals through the denoising process. This approach enables us to learn to generate domain-specific variance according to each subject, rather than separating subject noise through simple filtering or analysis.

\begin{figure*}[!t]
\centering
\subfloat[Original DDPM model]{\raisebox{0.18\height}{\includegraphics[width=0.32\textwidth]{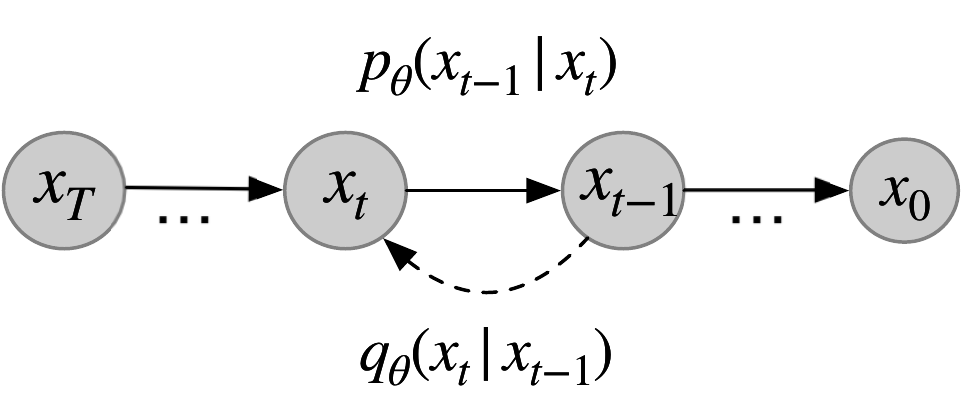}}\label{fig:math_org}}
\hfil
\subfloat[DS-DDPM model]{\includegraphics[width=0.44\textwidth]{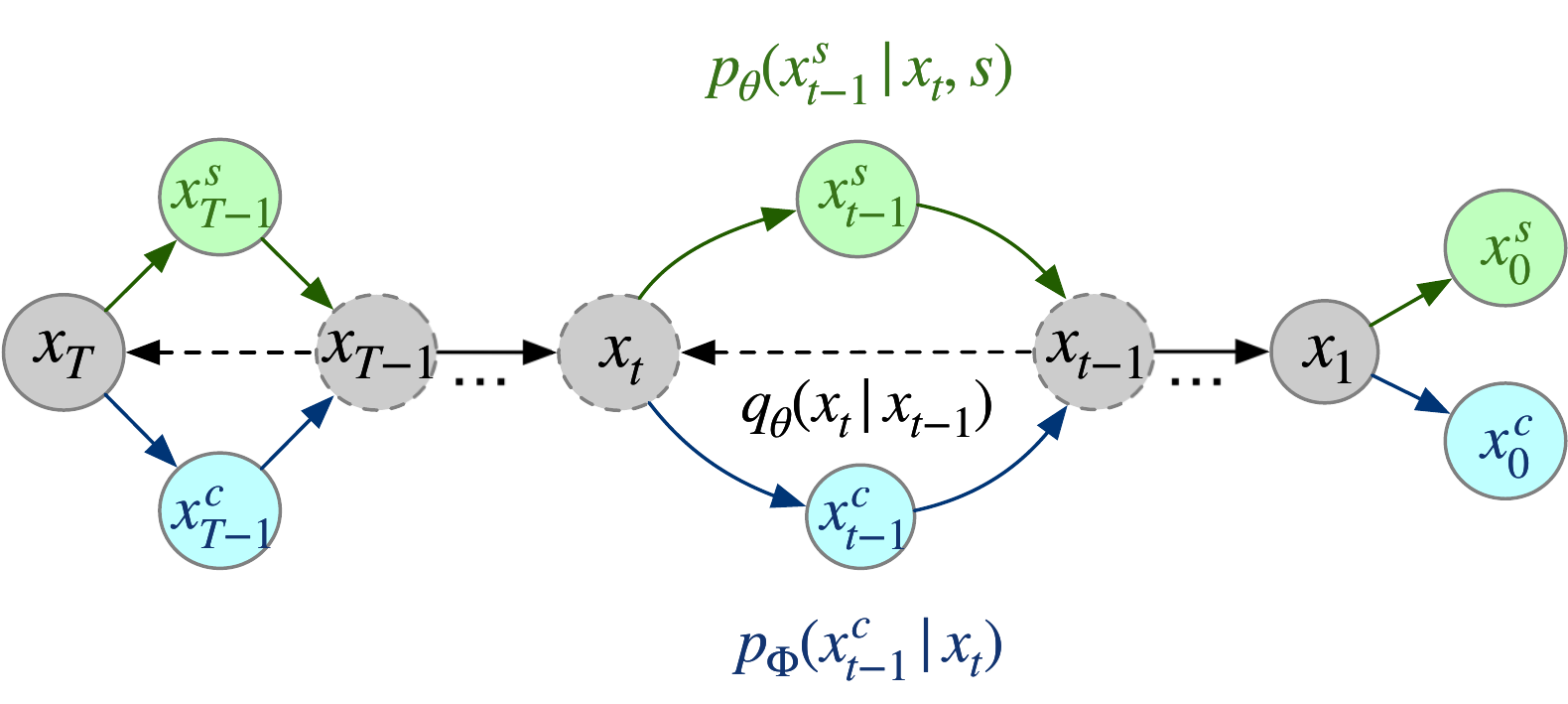}\label{fig:math_dsddpm}}
\caption{\label{fig:mathmodel} Illustration of probabilistic model graphs, where Figures~\ref{fig:math_org} and~\ref{fig:math_dsddpm} respectively denote the original model DDPM and the proposed DS-DDPM. DS-DDPM separate the subject-specific domain variance $\x^{s}_{t}$ (human artifacts), and the clean signal $\x^{c}_{t}$ at each denoising time step $t$. We constrain the summation of the separated $\x^{s}_{t}$ and $\x^{c}_{t}$ equal to diffusion result $\x_{t}$ at time step $t$, which denotes the two separated variance shares the same diffusion process. Here, the dotted line denotes the diffusion process, while the solid line denotes the denoising process.}
\end{figure*}

In this paper, we propose to revisit human artifact removal for EEG signals by modeling the removal process using domain-specific denoising diffusion probabilistic models (DS-DDPM). We adhere to the widely accepted assumption that recorded EEG signals comprise mixtures of clean samples and subject-specific human artifact noises.

We introduce a novel probabilistic model based on the diffusion-denoising generative process~\cite{nichol2021improved,choi2021ilvr,duan2023diffusiondepth}, which has recently gained prominence, particularly for image generation tasks~\cite{ramesh2021zero}. The proposed DS-DDPM incorporates artifact removal into every denoising step in the original model, as illustrated in Figure~\ref{fig:mathmodel}. At time step $t$, the diffusion result $\x_{t}$ is separated by two denoising models into domain variance $\x^{s}{t}$ (human artifacts) and the clean signal $\x^{c}{t}$. We represent the diffusion result as the summation of the two separated contents: $\x_{t} = \x^{c}{t} + \x^{s}{t}$, indicating that the two streams share the same diffusion process. This constraint naturally corresponds to our assumption that the noisy EEG signal can be decomposed into clean signals and domain-specific noise. A detailed mathematical problem definition is provided in Section~\ref{subsec:problemdef}.

We decompose the domain variance and clean signal into two mutually orthogonal spaces, as described in Section~\ref{subsec:domianspecific}. Additionally, an auxiliary cosine classifier is applied to disentangle variance spaces for subject-specific noises. Since no previous methods have explored DDPM models for EEG signals, Section~\ref{subsec:eeg-unet} offers technical details of the efficient structure of DDPM for EEG signals.

Comprehensive experimental results demonstrate that 1) the domain-specific separation is significant in a subject-domain-wise manner (Section~\ref{subsec:subcor}), and 2) domain-specific denoising can help enhance cross-subject classification performance (Section~\ref{subsec:cls}). We also present an ablation study discussing the contribution of each proposed component in Section~\ref{subsec:abl}.
The contributions of this paper can be summarized into three main aspects:

\begin{itemize}
    \item This work is the first to introduce denoising diffusion probabilistic models into EEG signals.
    \item DS-DDPM gives a novel approach to explicitly separate long-existing domain variance related to human subject difference through a domain-specific denoising process. 
    \item Comprehensive Experimental results suggest that the proposed DS-DDPM is efficient in domain-specific denoising for both relevance analysis and classification performance. 
\end{itemize}

\section{Related Works}
\subsection{Denoising Methods for EEG}

Artifact removal from EEG signals is a crucial step before extracting neural information for subsequent analyses. Artifacts, which may arise from imprecise recording systems, casual recording procedures, or human subjects themselves, can be categorized into intrinsic or extrinsic artifacts based on whether they originate from the human subject. Regardless of their source, these artifacts can lead to misleading results in brain-computer interface (BCI) applications \cite{mannan2018effect,zhou2023speech2eeg}. In particular, physiological artifacts caused by human subjects cannot be eliminated by merely applying bandwidth filtering or strict experimental procedures. Mainstream methods for physiological artifact removal include signal decomposition and artifact signal estimation, incorporating techniques such as regression \cite{al2018complexity,klados2011reg}, Blind Source Separation (BSS) \cite{sweeney2012artifact,somers2016removal,teixeira2006automatic,subasi2010eeg}, Empirical-mode Decomposition (EMD), and spectrum filtering methods \cite{james2004independent}.

Spectrum filtering methods, which involve transforming temporal signals into frequency components, enable the separation or extraction of signals. In the frequency domain, signals can be decomposed into distinct combinations of individual frequencies and amplitudes, thereby separating useful signals from artifacts. Common methods for spectrum filtering in EEG denoising include Wavelet Transform \cite{kumar2008removal,safieddine2012removal} and Fourier Transform \cite{behnam2007analyses,murugappan2013human}. However, these approaches typically require some prior knowledge about the characteristics of useful components and artifacts in the EEG signal, which may be limiting when applied to a specific EEG-related application without sufficient exploration and understanding of its EEG signal.

In contrast, BSS methods are unsupervised learning algorithms that learn an underlying linear transformation, transforming the source EEG signal to the observed noisy EEG signal without needing prior information and reference channels. Principal Component Analysis (PCA) \cite{berg1991dipole}, Independent Component Analysis (ICA) \cite{jung1998independent, vigario1997extraction, vigario2000independent, jung2000removal}, and Canonical Correlation Analysis (CCA) \cite{borga2002canonical,dong2015characterizing,de2006canonical} are some major BSS methods. Although these methods achieve high accuracy in artifact removal, they strongly assume that the source signals can be approximated by linear transformation from the noisy observation. We argue that this oversimplified transformation may not adequately represent the complex nature of EEG artifacts.

\subsection{Domain Separation}

EEG signals are well-known for their substantial inter-subject variance. Our method resembles domain separation and adaptation methods, as we aim to capture both the similarities in high-level EEG patterns and the differences in subject-related patterns using distinct models. Specifically, existing domain separation methods assume that signals can be decomposed into domain-specific and domain-invariant features. In our approach, the domain-invariant feature directly refers to the underlying mental state derived from EEG signals collected from human subjects, while the domain-specific feature corresponds to subject-specific features that manifest as unwanted artifacts in EEG analysis.

Early work on domain separation, such as Domain Separation Networks (DSN) \cite{bousmalis2016domain}, proposed an architecture to learn and separate both domain-specific and domain-invariant features by introducing additional network branches for each feature. By reducing the discrepancy between the domain-specific subspace and the class-specific feature space, their method can learn separable domain-specific information that is orthogonal to the domain-invariant feature space. Later, Adversarial Discriminative Domain Adaptation (ADDA) \cite{tzeng2017adversarial} employed a domain discriminator~\cite{duan2023cross} to allocate features obtained from different domains into a shared space. In this way, domain-specific information is not captured by the feature extractor for different domains.

More recent approaches proposed further separation of domain-specific and domain-invariant features by utilizing additional regularization terms that eliminate the dependency between domain-specific features and domain-invariant features. These terms include Entropy regularization \cite{zhao2020domain} and Variance Penalty \cite{heinze2021conditional,wang2017generalized}. Although these methods vary in network architectures and regularizations, the core idea behind domain separation can be summarized as eliminating the discrepancy between domain-specific features and domain-invariant features. To the best of our knowledge, our work is the first to establish a connection between domain separation and subject modeling in EEG processing.

\section{Methodology}
\label{sec:method}
Section~\ref{subsec:problemdef} provides the mathematical description of how we revisit the human artifact removal by formulating it as a domain-specific denoising diffusion probabilistic process. 
Section~\ref{subsec:domianspecific} provides technical details about how we constrain and separate the two variables through novel training loss.
Section~\ref{subsec:eeg-unet} illustrate how we design the model structure to fit the properties of EEG signals. 

\subsection{Definition of Domian-Sepcific Denoising}\label{subsec:problemdef}

The separation of domain-specific variance is decomposed into the denoising process of a denoising diffusion probabilistic model~\cite{nichol2021improved}. 
We define $\x_{0}$ as the original recorded EEG signals and $\x_t$ as the variable by adding Gaussian noise distribution by sequentially $t$ times iteration. 
Thus, we could continuously add noise into original $\x_{0}$ through a Markov process sampling variables $\{\x_0,\x_1,...\x_{t-1},\x_{t},...,\x_T\}$ until $\x_T$ becomes a normal noise distribution $p(\x_T)\sim \mathcal{N}(\x_T;0,I)$ as shown Figure~\ref{fig:mathmodel}. Here, the transition is also called \textit{diffusion process or forward process} as below.
\begin{align}
    q(\x_{1:T}|\x_0) &:= \prod^{T}_{t=1}q(\x_t|\x_{t-1}), \nonumber \\
    q(\x_t|\x_{t-1}) &= \mathcal{N}(\x_t; \sqrt{1-\beta_t} \x_{t-1}, \beta_{t}I)
\end{align}

where ${\beta_1, \beta_2,...\beta_{T}}$ is a fixed variance coefficient schedule. Following the Gaussian distribution assumption of DDPM, $\x_t$ could be further represented as the combination of $\x_0$ and sampled variance $\boldsymbol{\varepsilon}$.
\begin{align}\label{eq:diffq}
    q(\x_{t}|\x_0) &:= \mathcal{N}(\x_t;\sqrt{\Bar{\alpha}_t}\x_0, (1-\Bar{\alpha}_t)\I), \nonumber \\  \alpha_t &:= 1-\beta_t, \quad \Bar{\alpha}_t =  \prod^t_{i=1}\alpha_i
\end{align}
where $\alpha_t$ is also a fixed variance coefficient schedule corresponding to $\beta_t$. 
In practice, the representation of $\x_t$ could be obtained by extending the diffusion process defined in Equation~\ref{eq:diffq} as below.
\begin{align}\label{eq:diffusionx}
\x_t &= \alpha_t\x_{t-1} + \beta_t \boldsymbol{\varepsilon}_t = \alpha_t\left(\bar{\alpha}_{t-1}\x_0 + \bar{\beta}_{t-1}\bar{\boldsymbol{\varepsilon}}_{t-1}\right) + \beta_t \boldsymbol{\varepsilon}_t \nonumber \\
&=\bar{\alpha}_t\x_0 + \alpha_t\bar{\beta}_{t-1}\bar{\boldsymbol{\varepsilon}}_{t-1} + \beta_t \boldsymbol{\varepsilon}_t 
\end{align}
where $\boldsymbol{\varepsilon_t} \sim \mathcal{N}(0,\I)$ is a Gaussian distribution that represents the stochastic property of the diffusion process. It also gives a description of how to represent the diffusion result $\x_t$ by real sample $\x_0$ and given fixed variance scheduler $\alpha_t$ and $\beta_t$.  

Different from classical DDPMs, which model a direct \textit{reverse process or denoising process} as $q(\x_{t-1}|\x_{t})$, we decompose the denoising process for different subject domains $s$ by separating the process into two variables: $q(\x^s_{t-1}, \x^c_{t-1}|\x_{t})$. Here, $\x^s_{t-1}$ represents the separated domain variance for subject $s$, and $\x^c_{t-1}$ represents the corresponding clean signal generated at step $t$. We assume the signal is a mixture of subject noise and the clean signal, such that $\x_{t-1} = \x^s_{t-1} + \x^c_{t-1}$.
As the direct reversal of the diffusion process $q(\x^s_{t-1}, \x^c_{t-1}|\x_{t})$ is intractable~\cite{sohl2015deep}, we employ two separate functions to express the denoising process as follows.
\begin{align}
    p_{\theta}(\x^s_{t-1}|\x_t) &:= \mathcal{N}(\x^s_{t-1};\mu_{\theta}(\x_t,t,s), \sigma_t^2\I), \nonumber
    \\
    p_{\phi}(\x^c_{t-1}|\x_t) &:= \mathcal{N}(\x^c_{t-1};\mu_{\phi}(\x_t,t), \sigma_t^2\I)
\end{align}
Where $p_{\theta}(\x^s_{t-1}|\x_t)$ and $p_{\phi}(\x^c_{t-1}|\x_t)$ are the two separated denoising functions decomposed from $p(\x_{t-1}|\x_t) = p_{\theta}(\x^s_{t-1}|\x_t) + p_{\phi}(\x^c_{t-1}|\x_t)$. Here, $\sigma_{t}^2$ denotes the variance in transition. The core transitions $\mu_{\theta}(\x_t,t,s)$ and $\mu_{\phi}(\x_t,t)$ are learned by deep neural networks. We follow previous experimental settings~\cite{nichol2021improved,choi2021ilvr}, where $\sigma_{t}^2$ is directly set as $\beta_t$ or $\frac{1-\Bar{\alpha}_t}{1-\alpha_t}\beta_t$, which have shown similar results in previous experiments.
Thus, the variable $\x_t$ at time step $t$ could be expressed as the summation of domain variance, and the clean data follow the original DDPM conduction~\cite{nichol2021improved}.
\begin{align}
    \x_{t-1} &= \frac{1}{1-\alpha_t}(\x_t - \frac{1-\alpha_t}{\sqrt{1-\Bar{\alpha}_t}}\mu) + \sigma_t \z, \\
    \mu &= \mu_{\theta}(\x_t,t,s) + \mu_{\phi}(\x_t,t)
\end{align}
where the two learned transition deep model $\mu_{\theta}(\x_t,t,s)$ and $\mu_{\phi}(\x_t,t)$ share a same variance coefficient schedule corresponding to $\beta_t$. Given relation we defined above $\x_{t-1}=\x^s_{t-1}+\x^c_{t-1}$, we could further approximate the separated domain variance of subject $s$ as $\x^s_{t-1} = \frac{1}{1-\alpha_t}(\x_t - \frac{1-\alpha_t}{\sqrt{1-\Bar{\alpha}_t}} \mu_{\theta}(\x_t,t,s)) + \sigma_t \z$, and the separated clean signal as $\x^c_{t-1} = \frac{1}{1-\alpha_t}(\x_t - \frac{1-\alpha_t}{\sqrt{1-\Bar{\alpha}_t}} \mu_{\phi}(\x_t,t)) + \sigma_t \z$. It is noted that, after the denoising process sampled to $\x_1$, we could directly calculate the desired domain variance $\x^s_0$ of human subjected $s$ and clean data $\x^c_0$ according to the equations defined above. 

\subsection{Separate Domain Specific Variance by Constraint}\label{subsec:domianspecific} 
Under the domain-specific denoising process defined in Section~\ref{subsec:problemdef}, we separate the domain-specific variance by the combination of three constraints. 
1) Section~\ref{subsub:reserve} introduces how we constraint the summation separated clean signal and domain-specific variance could reconstruct the EEG signals by reverse process loss $\mathcal{L}_r$
2) Section~\ref{subsub:org} introduces orthogonal constraint between clean signal and domain-specific variance by orthogonal loss $\mathcal{L}_{o}$. 
3) Section~\ref{subsubsec:arcface} further formulate the domain variance space separable according to different human subjects by Arc-Marging loss $\mathcal{L}_{arc}$.

\subsubsection{Reverse Process}\label{subsub:reserve}

Basically, we model the diffusion results $\x_t$ at time step $t$ as the mixture of domain variance $\x^s_t$ and clean data $\x^c_t$ as defined in Section~\ref{subsec:problemdef} and Figure~\ref{fig:math_dsddpm}. We first discuss how to ensure the effectiveness of diffusion-denoising training.
Given the decomposition the of denoising results as  $\x_t = \x^s_t + \x^c_t$, generative training loss is granted by minimizing the distance ${\Vert \x_{t-1} -(\mu_{\theta}(\x_t,t,s) + \mu_{\phi}(\x_t,t))\Vert}^2$ between diffusion results $\x_t$ and denoising results. 
Follow the original conduction of DDPM~\cite{nichol2021improved}, 
Considering the definition in Section~\ref{subsec:problemdef} that the separated two denoising models share the same variance coefficient scheduler $\alpha$ and $\beta$ we could approximate the summation of the denoising process as.
\begin{align}\label{eq:denoisingmu}
    \mu_{\theta}(\x_t,t,s) + \mu_{\phi}(\x_t,t) = \nonumber \\ \frac{1}{\alpha_t}(\x_t - \beta_t\boldsymbol{\varepsilon}_{\theta}(x_t, t, s) - \beta_t\boldsymbol{\varepsilon}_{\phi}(x_t, t)) 
\end{align}
where $\theta$ and $\phi$ are respectively the training parameter for domain-specific denoising and content denoising. $\boldsymbol{\varepsilon}_{\theta}$ and $\boldsymbol{\varepsilon}_{\phi}$ denote the generated variance from these models. 
Thus by introducing Equation~\ref{eq:denoisingmu}, we could rewrite the distance to be minimized as in Equation~\ref{eq:redistance}. 
\begin{align}\label{eq:redistance}
    {\Vert \x_{t-1} -(\mu_{\theta}(\x_t,t,s) + \mu_{\phi}(\x_t,t))\Vert}^2 = \nonumber \\ \frac{\beta_t^2}{\alpha_t^2}{\Vert \boldsymbol{\varepsilon}_t - \boldsymbol{\varepsilon}_{\theta}(x_t, t, s) - \boldsymbol{\varepsilon}_{\phi}(x_t, t))\Vert}^2
\end{align}
where $\frac{\beta_t^2}{\alpha_t^2}$ is a loss coefficient which we use hyper parameter $\lambda_r$ to represent. Also by introducing the diffusion process to represent $\x_t$ defined in Equation~\ref{eq:diffusionx}, we could give the training loss $\mathcal{L}_r$ for the reverse process in Equation~\ref{eq:reverseloss}.

\begin{align}
\label{eq:reverseloss}
    \mathcal{L}_r = & \lambda_r \Vert \boldsymbol{\varepsilon}_t - \boldsymbol{\varepsilon}_{\theta}(\bar{\alpha}_t\x_0 + \alpha_t\bar{\beta}_{t-1}\bar{\boldsymbol{\varepsilon}}_{t-1} + \beta_t \boldsymbol{\varepsilon}_t, t, s) \nonumber \\ & - \boldsymbol{\varepsilon}_{\phi}(\bar{\alpha}_t\x_0 + \alpha_t\bar{\beta}_{t-1}\bar{\boldsymbol{\varepsilon}}_{t-1} + \beta_t \boldsymbol{\varepsilon}_t, t))\Vert^2 
\end{align}

The reverse process training loss $\mathcal{L}_r$ could be minimized by given recorded EEG signals $\x_0$, fixed variance scheduler $\{ \alpha_{1:T}, \beta_{1:T}\}$ and standard Gaussian distribution $\boldsymbol{\varepsilon}_t$ sampled at each time step.

\subsubsection{Separate Subject Domain Variance Apart}\label{subsub:org}

\begin{figure}[hbpt]
\centering
\vspace{-10pt}
\includegraphics[width=0.28\textwidth,height=0.2\textwidth]{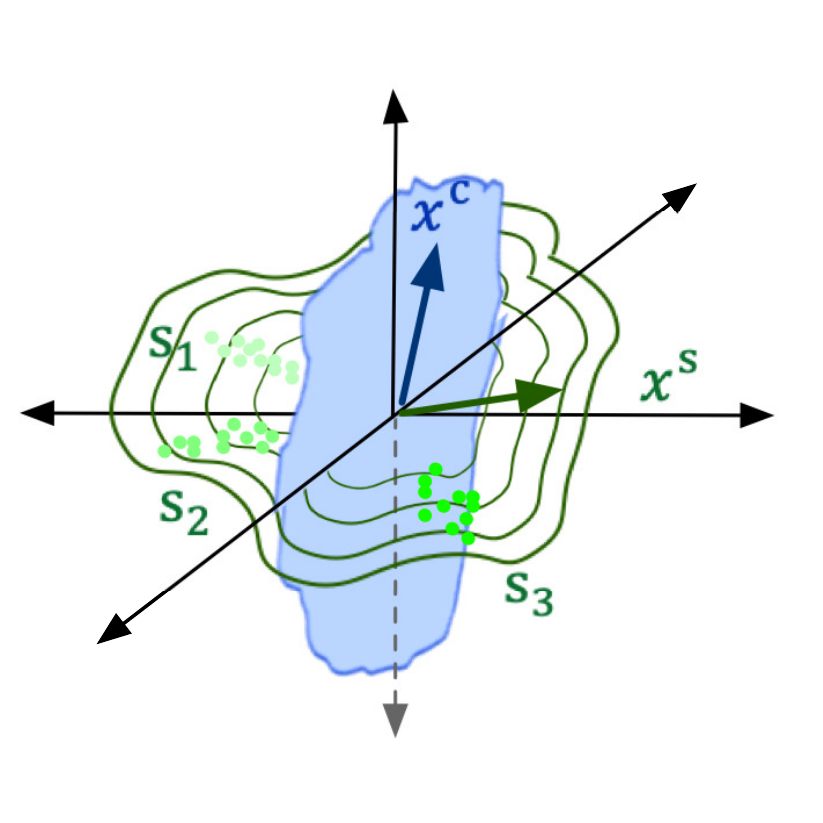}
\vspace{-10pt}
\caption{\label{fig:domainspace} The orthogonal domain variance separation of DS-DDPM.}
\end{figure}

The training of the reverse process only ensures the summation of the generated $\x^s_0$ and $\x^c_0$ is the recorded EEG signal $\x_0$. Yet, we impose two constraints to separate the domain variance. 
First, we formulate the generated $\x^s_0$ and $\x^c_0$ at each time step into two orthogonal spaces as shown in Figure~\ref{fig:domainspace}, where the blue space contains the signal distribution of the clean EEG signals, green space contains the distribution of the subject-specific domain variance. 

The orthogonal property is granted by introducing a constraint loss defined in Equation~\ref{eq:orgloss}. 
\begin{equation}\label{eq:orgloss}
    \mathcal{L}_{o} = \lambda_{o}\left\Vert\left(\mu_{\theta}(\x_t,t,s)^{\top}\mu_{\phi}(\x_t,t)-\boldsymbol{I}\right)\otimes (1 - \boldsymbol{I})\right\Vert^2\ 
\end{equation}
where $\mu_{\theta}(\x_t,t,s)$ and $\mu_{\phi}(\x_t,t)$ are the denoising process defined in Equation~\ref{eq:denoisingmu}. According to this representation, we make an approximation to directly optimize the model output $\boldsymbol{\varepsilon}_{\theta}(x_t, t, s)^{\top}\boldsymbol{\varepsilon}_{\phi}(x_t, t)$ instead, the accordingly adjustment of the coefficient $\lambda_{o}$ regularize the value scale of $\mathcal{L}_{o}$.

\subsubsection{Formulate Subject Domain Variance to Human Subjects} \label{subsubsec:arcface}

In order to carry out domain-specific denoising pertaining to a particular human subject $s$, the DS-DDPM model also enforces subject-wise separability of the domain variance. This constraint is achieved by introducing an additional subject classifier to supervise the conditional denoising process $\mu_{\theta}(\x_t, t, s)$. The domain variance generated for subject $s$ is expected to predict the subject label $s$ based on the domain variance, by introducing a conditional probabilistic model $p(s | \x^s_t)$.

Under our assumption, we design the clean data variance to be entirely independent of the subject information. Consequently, we can eliminate all terms in the equations previously defined for representing the classification model $p(s | \x^s_t)$. We employ a simple EEGNet~\cite{lawhern2018eegnet} classifier to extract the feature $\bar{\x}^{s_i}t$ from the denoising output $\x^s_t$, with the transition represented as $\W{\theta}$.

We propose formulating the decision boundary for predicting subject $s$ in the cosine space, where the traditional SoftMax~\cite{liu2016large} loss is replaced by the Additive Angular Margin classification (Arc-Margin) loss~\cite{deng2019arcface}. The Arc-Margin loss transforms the classification logits as $\W_{\theta,s_j}^{\top} \bar{\x}^{s_i}t = \Vert \W{\theta,s_j}\Vert \Vert \bar{\x}^{s_i}t\Vert \cos \theta{s_j}$, where $\theta_{s_j}$ is the angle between the weight $\W_{\theta,s_j}$ and the feature $\bar{\x}^{s_i}t$ representing subject ${s_i}$. The individual weight and features are fixed by $L_2$ normalization. Here, we let $\Vert \W{\theta,s_j}\Vert =1 $, which signifies that the weights make predictions solely dependent on the angle between the feature and the weight. We also let $\Vert \bar{\x}^{s_i}_t\Vert = {r}$, representing that the learned embedding features are distributed on a hyper-sphere with a radius of ${r}$. As the feature $\bar{\x}^{s_i}t$ is calculated by $\mu{\theta}(\x_t, t, s)$ through a series of classification transformations $\omega$, $\bar{\x}^{s_i}_t$ can be approximated by eliminating irrelevant terms in Equation~\ref{eq:denoisingmu} as follows.

\begin{equation}\label{eq:cosinedef}
 \resizebox{0.435\textwidth}{!}{$
    \cos \theta_{s_j} 
    = \frac{\W_{\theta,s_j}^{\top}\omega(\boldsymbol{\varepsilon}_{\theta}(\bar{\alpha}_t\x_0 + \alpha_t\bar{\beta}_{t-1}\bar{\boldsymbol{\varepsilon}}_{t-1} + \beta_t \boldsymbol{\varepsilon}_t, t, s))}{\Vert \W_{\theta,s_j}^{\top}\Vert \Vert \omega(\boldsymbol{\varepsilon}_{\theta}(\bar{\alpha}_t\x_0 + \alpha_t\bar{\beta}_{t-1}\bar{\boldsymbol{\varepsilon}}_{t-1} + \beta_t \boldsymbol{\varepsilon}_t, t, s)) \Vert} $
   }
\end{equation}

An additive angular margin penalty $m$ is added between $\bar{\x}^{s_i}_t$ and $\W_{\theta,s_j}$ to simultaneously enhance the intra-class compactness and inter-class discrepancy~\cite{deng2019arcface}. Also by giving the coefficient $\lambda_{arc}$ for subject classification, we could formulate the Arc-Margin loss $\mathcal{L}_{arc}$ could be defined in Equation~\ref{eq:arcmargin}.

\begin{equation}\label{eq:arcmargin}
\resizebox{0.435\textwidth}{!}{$
    \mathcal{L}_{arc} =  -\lambda_{arc} \frac{1}{N}\sum^{N}_{i=1}\log\frac{e^{r\left(\cos\left(\theta_{y^i_{s}} + m\right)\right)}}{e^{r\left(\cos\left(\theta_{y^i_{s}} + m\right)\right)} + \sum^{n}_{s_j=1, s_j \neq y^i_{s}}e^{r\cos\theta_{s_j}}} $
    }
\end{equation}

where $y^i_{s}$ denotes the target subject label. By introducing $\mathcal{L}_{arc}$, we could regulate the domain variance space separable by angle according to different subject $s_i$. 
Since we already regulate the clean data space and the subject space orthogonal, the mixture of different subjects will be distributed evenly according to different subjects $s_i$ as well, which increases the interoperability of DS-DDPM.

\subsubsection{Temporal Difference Constraint} \label{subsubsec:temporalconstraint}

\begin{figure}[t]
\centering
\includegraphics[width=0.49\textwidth]{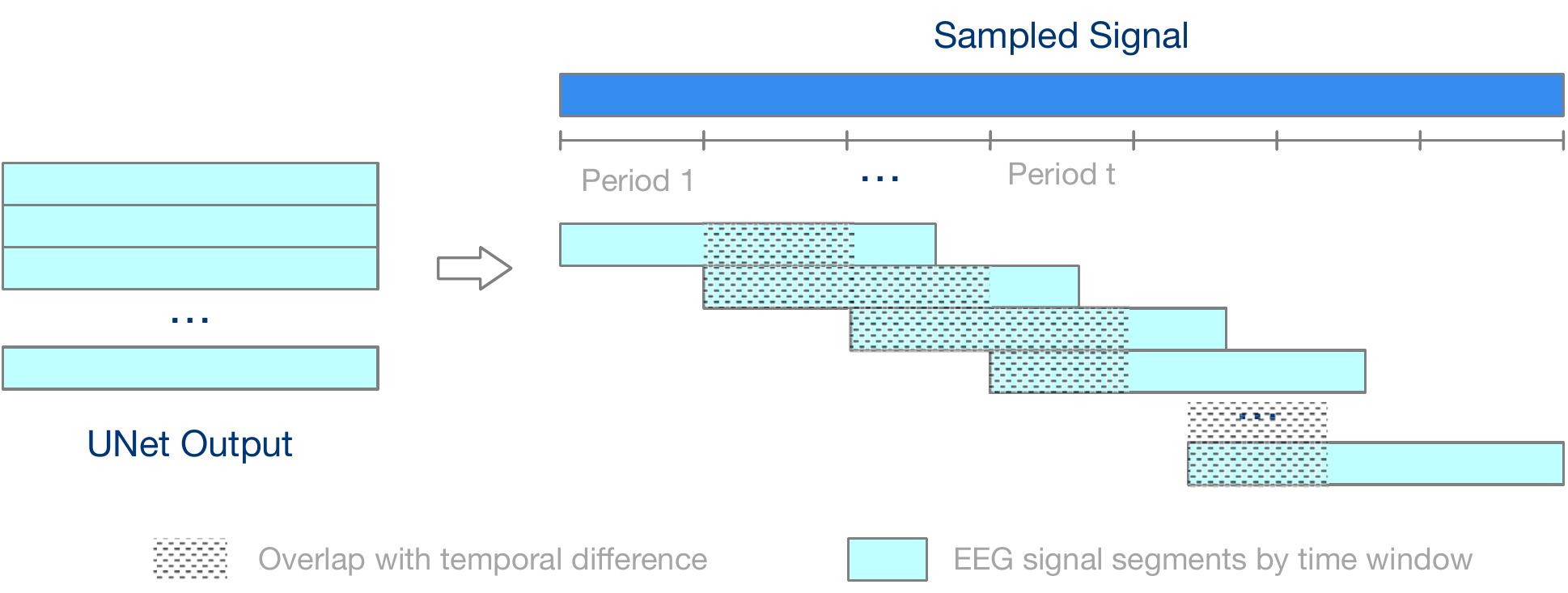}
\caption{\label{fig:overlap} Illustration of generative temporal difference in generative process. }
\end{figure}

As described in the methodology, we create the network inputs by stacking segments with overlapping regions in both temporal and spatial dimensions, as illustrated in Figure~\ref{fig:overlap}. Consequently, the results generated by the EEG-UNet model also contain overlaps. When performing the denoising process, we expect the same time points to have deterministic outputs, which means we aim to minimize the temporal difference in overlaps. To address this, we constrain the temporal difference by minimizing $\mathcal{L}_{td}$, as defined in Equation~\ref{eq:ltd}.

\begin{equation}\label{eq:ltd}
\mathcal{L}{td} = \frac{1}{n} \sum{i=1}^{n} \text{Overlap}(\y_{t-1} - {\y}_t)^2, \quad \y_t \in \x_0
\end{equation}

In this equation, we calculate the Mean Squared Error (MSE) between the sequence of sampled segments $\y_t$ generated by dividing $\x_0$ into stackings. The constraint is achieved by relaxing it into a regularization term and incorporating it into the main loss function.

\subsection{Model Structure}\label{subsec:eeg-unet}

\begin{figure*}[hbtp]{
\centering
\includegraphics[width=0.94\textwidth]{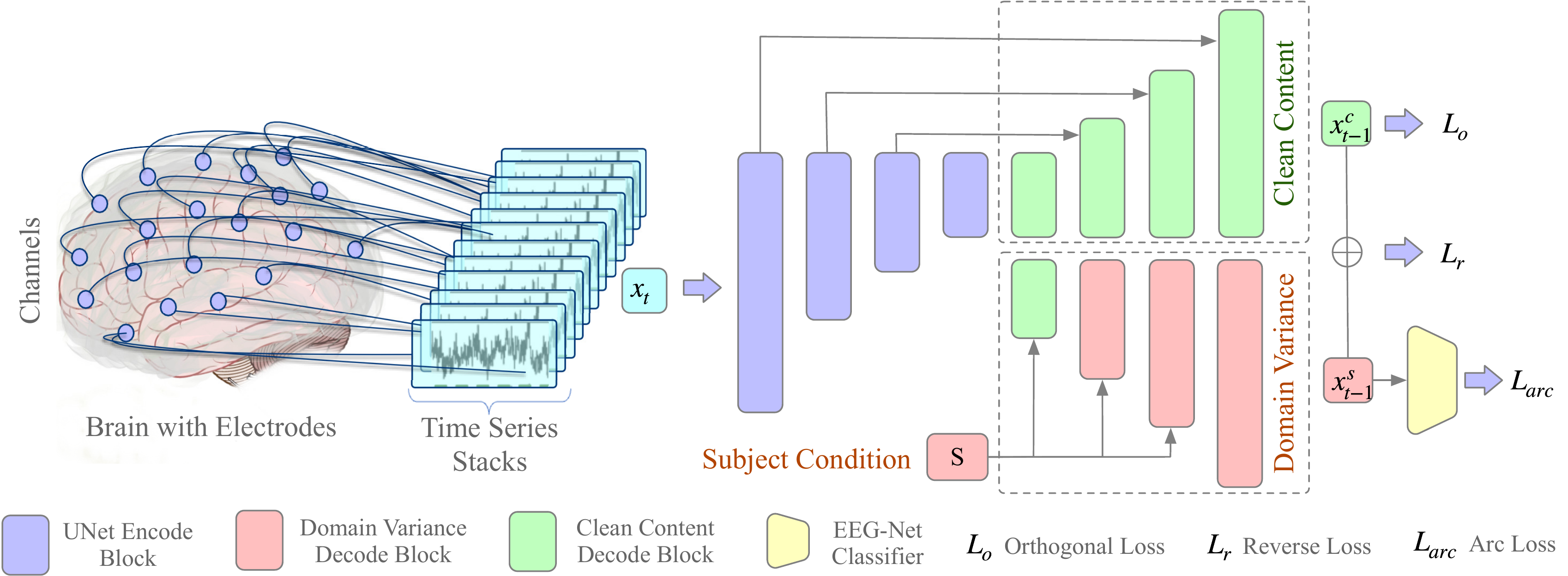}
\caption{\label{fig:eegnet}The overall model structure of DS-DDPM, where the recorded EEG signals are sliced by time window 224, step 75 and stacked 
into shape $bs\times channel \times window \times stacks$ as $\x_0$. The modified UNet-EEG structure receives diffusion results $\x_t$ and respectively generates domain variance $\x_{t-1}^s$ and the clean data $\x_{t-1}^c$ by two streams. The generated $\x_{t-1}^s$ and $\x_{t-1}^c$ are simultaneously supervised by reverse loss $\mathcal{L}_{r}$, orthogonal loss $\mathcal{L}_{o}$ and Arc-Margin loss $\mathcal{L}_{r}$. }
}
\end{figure*}

This section introduces how we design domain-specific generative model structure and how we train the denoising model by combining the loss constrained proposed in Section~\ref{subsec:domianspecific}. 
For model structure, we follow the common UNet~\cite{huang2020unet} structure and modified it to fit long-time series signals (UNet-EEG).
The overall model structure is shown in Figure~\ref{fig:eegnet}, where we modified the UNet~\cite{huang2020unet} structure into two generative streams.

The UNet-EEG model receives diffusion results $\x_t$ and respectively generates domain variance $\x_{t-1}^s$ and the clean content $\x_{t-1}^c$.
Here for the clean content stream, the UNet-EEG sequentially perform three times down-sample and up-sample operation with the residual fusion within each feature map scale, which could enhance the model's ability to maintain time sequential relations. 
Similar to the original DDPM, time label $t$ is tokenized into time embedding and fed into each layer of the UNet-EEG to improve the generative ability according to each time step $t$. 
For the domain variance separation stream, the subject condition $s$ is tokenized into embedding and fused with UNet-EEG at the mid-layer by multi-head attention layers. Here, we take the subject token as the query and the original feature map as the key and value for the attention layer.

\textbf{Training and Sampling Procedure:} In light of the structure defined above, we outline a comprehensive training procedure as presented in Algorithm~\ref{alg:training}. To expedite convergence, we pre-train the subject classifier $\omega$ (EEGNet) by sampling ${\x_0, s} \sim q(\x_0, s)$ pairs from the recorded EEG signals of a group of human subjects. As EEG signals inherently contain subject labels (indicating the human source of the EEG signal), training of $\omega$ is feasible. This approach helps the optimizer focus on the parameters $\theta$ and $\phi$ for the generative model.

For the entire training procedure, the time step $t$ is sampled uniformly from ${1, 2, ..., T}$, and the variance scheduler ${\alpha, \beta}$ is sampled and fixed. At each iteration, the algorithm samples $\boldsymbol{\varepsilon} \sim \mathcal{N}(\mathbf{0}, \I)$ at every time step. During each iteration, the generative model for clean data and subject-specific domain variance is computed, taking into account the subject label $s$, time step $t$, real EEG signal $\x_0$, and the sampled variables mentioned above. The parameters $\theta$ and $\phi$ of the generative model are optimized by minimizing the weighted combination of the loss functions proposed in Section~\ref{subsec:domianspecific}.

\begin{figure}[hbpt]
\begin{minipage}[t]{0.45\textwidth}
\begin{algorithm}[H]
  \caption{Training} \label{alg:training}
  \small
  \begin{algorithmic}[1]
    \WHILE{Not Converge}
      \STATE $\x_0, s \sim q(\x_0, s)$, {Sample $\x_0$ from Subjects $\{s\}$}
      \STATE Pretrain subject classifier $\omega$ given sample label pairs $\{\x_0, s\} $
      \STATE $t \sim \mathrm{Uniform}(\{1, \dotsc, T\})$
      \STATE $\boldsymbol{\varepsilon} \sim\mathcal{N}(\mathbf{0},\I)$
      \STATE Calculate, $\boldsymbol{\varepsilon}_{\theta}(x_t, t, s)$ , $\boldsymbol{\varepsilon}_{\phi}(x_t, t)$,  $\omega(\boldsymbol{\varepsilon}_{\theta}(x_t, t, s))$ with model parameters $\theta$ and $\phi$. 
      \STATE Take gradient descent step optimize $\theta$ and $\phi$ on 
      \STATE Loss $\mathcal{L}= \lambda_r \mathcal{L}_{r} + \lambda_o \mathcal{L}_{o} + \lambda_{arc} \mathcal{L}_{arc} + \lambda_{td} \mathcal{L}_{td}$
    \ENDWHILE
  \end{algorithmic}
\end{algorithm}
\end{minipage}
\hfill
\begin{minipage}[t]{0.45\textwidth}
\begin{algorithm}[H]
  \caption{Sampling} \label{alg:sampling}
  \small
  \begin{algorithmic}[1]
    \STATE $\x_T \sim \mathcal{N}(\mathbf{0}, \I)$
    \FOR{$t=T, \dotsc, 1$}
      \STATE $\z \sim \mathcal{N}(\mathbf{0}, \I)$ if $t > 1$, else $\z = \mathbf{0}$
      \STATE $\x^s_{t-1} = \frac{1}{\sqrt{\alpha_t}}\left( \x_t - \frac{1-\alpha_t}{\sqrt{1-\bar\alpha_t}} \boldsymbol{\varepsilon}_\theta(\x_t, t, s) \right) + \sigma_t \z$
      \STATE $\x^c_{t-1} = \frac{1}{\sqrt{\alpha_t}}\left( \x_t - \frac{1-\alpha_t}{\sqrt{1-\bar\alpha_t}} \boldsymbol{\varepsilon}_\phi(\x_t, t) \right) + \sigma_t \z$
    \ENDFOR
    \STATE \textbf{return} $\x^s_0$, $\x^c_0$
  \end{algorithmic}
\end{algorithm}
\end{minipage}
\end{figure}

After training the generative model, we can generate subject-specific domain variances based on the probabilistic model described in Section~\ref{subsec:problemdef}. Algorithm~\ref{alg:sampling} presents a comprehensive procedure for sampling domain variances according to subject $s$ and generating unconditional EEG signals from pure noise.

A straightforward approach for denoising a given signal involves sampling a subject-specific domain variance $\x^s_0$ and directly computing $\x_0 - \x^s_0$ to obtain a cleaner signal. However, this method does not consider the noise present in the given signal. Furthermore, the iterative denoising process can be time-consuming. In real-life scenarios, it is more practical to perform denoising based not only on subject $s$ but also on the signal itself.

Taking into account the denoising iterative process $\x_T,...,\x_t,\x_{t-1},...\x_0$, the denoising model incrementally removes noise at each step. Since recorded EEG signals inherently contain noise, it is reasonable to directly perform denoising by assuming the recorded signal as $\x_{1/2}$, an intermediate state in the denoising process. As a result, given a recorded raw signal $\x_{raw}$, we can directly separate the domain variance and clean content by computing: $\x^s_{0} = \frac{1}{\sqrt{\alpha_1}}\left( \x_{raw} - \frac{1-\alpha_1}{\sqrt{1-\bar\alpha_1}} \boldsymbol{\varepsilon}\theta(\x{raw}, 1, s) \right) + \sigma_1 \z$
and $\x^c_{0} = \frac{1}{\sqrt{\alpha_1}}\left( \x_{raw} - \frac{1-\alpha_1}{\sqrt{1-\bar\alpha_1}} \boldsymbol{\varepsilon}\theta(\x{raw}, 1) \right) + \sigma_1 \z$.

This assumption allows for replacing the iterative sampling process with a single inference step. Experimental results demonstrate that this approach effectively improves the cross-subject classification task.

\section{Experiments}

\label{sec:exp}

In this section, we present comprehensive experiments to demonstrate the efficiency of the proposed Domain-Specific Denoising. We analyze the correlation in Section~\ref{subsec:subcor} by 1) performing correlation coefficient calculations. Furthermore, we conduct experiments on the cross-subject classification task in Section~\ref{subsec:cls} to illustrate our method's efficiency in classification tasks. An additional ablation study is conducted in Section~\ref{subsec:abl} to discuss the effectiveness of each component.

\subsection{Experimental Setup}
We primarily conduct our experiments on the BCI-Competition-IV dataset~\cite{tangermann2012review}, which is widely used for validating Motor Imagery classification tasks. The dataset is collected under the widely-used 10-20 system, comprising 22 EEG channels and 3 EOG channels. The dataset contains EEG and EOG signals with a sampling frequency of 250 Hz from nine subjects. Each subject was required to perform four classes of motor imagery (left hand, right hand, feet, and tongue) while recording brain dynamics.

We use only the 22-channel EEG signals, preprocessing them into a shape of $22 \times 750$, where $750$ is the time sequence length with a 250 Hz sampling rate for 3 seconds. As mentioned in Section~\ref{subsec:eeg-unet}, we use a time window to slide along the time sequence and slice the time sequence into segments with overlaps. By using a time window size of 224 and stride size of 75, the data is processed into a shape of $bs \times 22 \times 224 \times 8$ and fed into the UNet-EEG model. For the UNet-EEG model, we sequentially perform three down-sampling and up-sampling blocks for the clean content stream and one down-sampling and up-sampling block for the domain variance stream. This slim structure for the domain variance stream prevents severe overfitting, making the training process more stable.

For the subject classifier, we use the original EEGNet structure with a modified input size of $bs \times 22 \times 224 \times 8$. For the Arc-Margin $\mathcal{L}_{arc}$ hyperparameters, we follow previous explorations on human face recognition, where we set the radius $r=30$ and the margin as $m=0.5$. We use the ADAM optimizer~\cite{zhang2018improved} with default hyperparameters. According to our experiments, we found that a batch size of $64$ stabilizes the training process more than smaller numbers. The DS-DDPM is implemented based on PyTorch, and we open-source our codes\footnote{\href{https://github.com/duanyiqun/DS-DDPM}{https://github.com/duanyiqun/DS-DDPM}.} to the community. 

\begin{table*}[hbpt]

\footnotesize
\setlength{\tabcolsep}{2mm}{}
\caption{\label{tb:crosssubjectcls} Cross-subject classification performance on BCI-IV dataset, where each column denotes a single model trained from one single subject and single denoising method, each row denotes which subject is selected for the training set.  \textbf{M} denotes the mean accuracy of each model.  }
\resizebox{0.99\textwidth}{!}{
\centering
\begin{tabular}{llllllllllllllllllll}
\toprule
                          & \multicolumn{9}{c}{Train with ICA Denoising (Acc. $\%$)}                                                                                                                                                                                                                                                  &  & \multicolumn{9}{c}{Train with DS-DDPM Denoising (Acc. $\%$)}                                                                                                                                                                                                                                              \\ \cline{2-10} \cline{12-20} 
                          & s1                            & s2                            & s3                            & s4                            & s5                            & s6                            & s7                            & s8                            & s9                            &  & s1                            & s2                            & s3                            & s4                            & s5                            & s6                            & s7                            & s8                            & s9                            \\ \cline{1-10} \cline{12-20} 
s1                        & \cellcolor[HTML]{EFEFEF}89.29 & 39.29                         & 25.93                         & 50.00                         & 40.74                         & 27.27                         & 62.96                         & 29.63                         & 36.00                         &  & \cellcolor[HTML]{EFEFEF}85.59 & 46.43                         & 46.43                         & 52.00                         & 37.04                         & 31.82                         & 68.22                         & 37.04                         & 38.46                         \\
s2                        & 82.61                         & \cellcolor[HTML]{EFEFEF}92.86 & 33.33                         & 41.67                         & 33.33                         & 31.82                         & 33.33                         & 37.04                         & 36.00                         &  & 85.71                         & \cellcolor[HTML]{EFEFEF}90.01 & 34.28                         & 44.00                         & 44.44                         & 37.27                         & 32.14                         & 44.18                         & 35.77                         \\
s3                        & 53.57                         & 85.71                         & \cellcolor[HTML]{EFEFEF}81.48 & 58.33                         & 33.33                         & 54.55                         & 44.44                         & 33.33                         & 48.00                         &  & 60.71                         & 92.86                         & \cellcolor[HTML]{EFEFEF}82.11 & 58.00                         & 58.15                         & 55.00                         & 49.29                         & 51.85                         & 44.62                         \\
s4                        & 53.57                         & 46.43                         & 81.48                         & \cellcolor[HTML]{EFEFEF}91.67 & 33.33                         & 54.55                         & 37.04                         & 37.04                         & 44.00                         &  & 52.86                         & 47.21                         & 78.57                         & \cellcolor[HTML]{EFEFEF}88.45 & 48.15                         & 50.00                         & 38.57                         & 40.74                         & 46.15                         \\
s5                        & 42.86                         & 35.71                         & 37.04                         & 91.67                         & \cellcolor[HTML]{EFEFEF}88.89 & 54.55                         & 40.74                         & 33.33                         & 40.00                         &  & 45.86                         & 35.71                         & 37.04                         & 91.67                         & \cellcolor[HTML]{EFEFEF}85.47 & 52.00                         & 42.81                         & 36.12                         & 46.15                         \\
s6                        & 50.00                         & 39.29                         & 37.04                         & 50.00                         & 77.78                         & \cellcolor[HTML]{EFEFEF}90.91 & 44.44                         & 33.33                         & 44.00                         &  & 52.14                         & 35.00                         & 46.43                         & 44.44                         & 79.12                         & \cellcolor[HTML]{EFEFEF}88.00 & 49.15                         & 33.33                         & 50.00                         \\
s7                        & 57.14                         & 50.00                         & 37.04                         & 41.67                         & 55.56                         & 90.91                         & \cellcolor[HTML]{EFEFEF}74.07 & 33.33                         & 40.00                         &  & 55.71                         & 50.00                         & 42.16                         & 52.00                         & 55.56                         & 90.91                         & \cellcolor[HTML]{EFEFEF}76.12 & 43.10                         & 47.28                         \\
s8                        & 39.29                         & 39.29                         & 44.44                         & 50.00                         & 59.26                         & 45.45                         & 66.67                         & \cellcolor[HTML]{EFEFEF}74.07 & 44.00                         &  & 39.29                         & 39.29                         & 44.44                         & 47.25                         & 59.26                         & 45.45                         & 70.37                         & \cellcolor[HTML]{EFEFEF}78.88 & 44.00                         \\
s9                        & 32.14                         & 32.14                         & 37.04                         & 45.83                         & 44.44                         & 45.45                         & 33.33                         & 74.07                         & \cellcolor[HTML]{EFEFEF}84.00 &  & 25.15                         & 35.71                         & 32.14                         & 38.20                         & 44.44                         & 45.45                         & 34.07                         & 73.08                         & \cellcolor[HTML]{EFEFEF}85.15 \\ \cline{1-10} \cline{12-20} 
\cellcolor[HTML]{EFEFEF}M & 55.61                         & 51.19                         & 46.09                         & \textbf{57.87}                & 51.85                         & \textbf{55.05}                & 48.56                         & 42.80                         & 46.22                         &  & \textbf{55.89}                & \textbf{52.47}                & \textbf{49.29}                & 57.33                         & \textbf{56.85}                & 54.48                         & \textbf{51.19}                & \textbf{48.70}                & \textbf{48.62}                \\ \bottomrule
\end{tabular}
}
\end{table*}

\begin{table}[hbpt]
\caption{\label{tb:corr}Subject-wise correlation analysis between real EEG samples and DS-DDPM generative samples. 
}
\setlength{\tabcolsep}{2mm}{}
\centering
\resizebox{0.479\textwidth}{!}{
\begin{tabular}{llllllllll}
\toprule
\multicolumn{10}{c}{Subject Correlation Coefficient of EEG BCI-IV EEG Signals}  \\ \hline
   & s1              & s2              & s3              & s4              & s5              & s6              & s7              & s8              & s9              \\ \hline
s1 & \textbf{0.101} & {\ul 0.048}    & 0.055          & {\ul 0.059}    & {\ul 0.045}    & 0.051          & 0.050          & 0.042          & 0.043          \\
s2 & 0.048          & \textbf{0.074} & 0.040          & 0.038          & 0.033          & 0.041          & 0.045          & 0.037          & 0.038          \\
s3 & 0.055          & 0.040          & \textbf{0.080} & 0.048          & 0.037          & 0.043          & 0.048          & 0.037          & 0.039          \\
s4 & {\ul 0.059}    & 0.038          & 0.048          & \textbf{0.088} & 0.041          & 0.042          & {\ul 0.053}    & {\ul 0.043}    & 0.044          \\
s5 & 0.045          & 0.033          & 0.037          & 0.041          & \textbf{0.058} & 0.034          & 0.039          & 0.031          & 0.034          \\
s6 & 0.051          & 0.041          & 0.043          & 0.042          & 0.034          & \textbf{0.071} & 0.044          & 0.034          & 0.038          \\
s7 & 0.050          & 0.045          & {\ul 0.548}    & 0.053          & 0.039          & 0.044          & \textbf{0.093} & 0.042          & {\ul 0.050}    \\
s8 & 0.042          & 0.037          & 0.037          & 0.043          & 0.031          & {\ul 0.054}    & 0.042          & \textbf{0.069} & 0.032          \\
s9 & 0.050          & 0.038          & 0.039          & 0.041          & 0.034          & 0.038          & 0.044          & 0.032          & \textbf{0.066} \\ \hline

\multicolumn{10}{c}{Subject Correlation Coefficient Between DS-DDPM Sampled Signal and BCI-IV}                                                                                                                           \\ \hline
s1 & \textbf{0.109} & {\ul 0.071}    & 0.077          & {\ul 0.080}    & 0.067          & {\ul 0.073}    & 0.072          & 0.063          & {\ul 0.070}    \\
s2 & 0.072          & \textbf{0.090} & 0.067          & 0.065          & 0.058          & 0.066          & 0.068          & 0.060          & 0.061          \\
s3 & 0.076          & 0.065          & \textbf{0.096} & 0.072          & 0.061          & 0.068          & 0.070          & 0.060          & 0.063          \\
s4 & 0.079          & 0.064          & 0.073          & \textbf{0.100} & 0.064          & 0.068          & {\ul 0.074}    & 0.064          & 0.064          \\
s5 & 0.069          & 0.060          & 0.065          & 0.067          & \textbf{0.076} & 0.062          & 0.063          & 0.055          & 0.059          \\
s6 & {\ul 0.083}    & 0.066          & 0.069          & 0.067          & 0.059          & \textbf{0.088} & 0.068          & 0.057          & 0.061          \\
s7 & 0.073          & 0.069          & {\ul 0.079}    & 0.075          & {\ul 0.070}    & 0.069          & \textbf{0.103} & {\ul 0.066}    & 0.066          \\
s8 & 0.067          & 0.064          & 0.065          & 0.069          & 0.057          & 0.062          & 0.066          & \textbf{0.082} & 0.057          \\
s9 & 0.072          & 0.064          & 0.067          & 0.067          & 0.059          & 0.065          & 0.067          & 0.056          & \textbf{0.082} \\ \hline
\end{tabular}
}
\end{table}

\subsection{Domain Specific Variance Analysis}\label{subsec:subcor}

\subsubsection{Correlation analysis} 
In order to analyze the generative quality of the domain variance generative tasks, we conduct correlation analysis on both real signals and generative signals, as shown in Table~\ref{tb:corr}. 
The upper part reports the correlation coefficient matrix on the real EEG signal distribution from BCI-Competition-IV 2a datasets. 
It shows that the coefficient between the same subjects is significantly higher than cross subjects' coefficient, where the diagonal values of the matrix are significantly larger than others. 
By giving the subject label to the DS-DDPM model, we could also generate real EEG signals of each subject. 
In order to illustrate the domain variance separation according to subject $s$, we analysis the coefficient between real samples and DS-DDPM-generated samples. 
If the generated signal distribution of subject $s$ has higher than the real signal distribution of the same human subject $s$, we think is significant to illustrate our efficiency. 
The results are reported in the lower part of Table~\ref{tb:corr}, where the correlation coefficient properties are similar to the real samples. 
These results indicate the efficiency of DS-DDPM. 
We could also observe that the generated signals have higher correlation coefficients with unrelated subjects compared to real signals, where the value of generated signals varies between $0.031\sim 0.059$. In comparison, the value of generated signals varies between $0.069\sim0.083$.
We argue that this phenomenon is rational as the real sample is actually the upper bound of the current regression-based domain separation methods.

\subsection{Cross-Subject Classification Performance}\label{subsec:cls}

To further verify the effectiveness of DS-DDPM in removing human artifacts, we conduct a cross-subject classification task and compare its performance with the widely used ICA method~\cite{subasi2010eeg}. In practice, we train MI classifiers for each subject individually using only their EEG signals and then test the classification accuracy on EEG signals from all other subjects. The results are reported in Table~\ref{tb:crosssubjectcls}.

We observe that by introducing DS-DDPM and utilizing the separated invariant features for classification tasks, the proposed model outperforms the previous method for subjects $1, 3, 4, 5, 7, 8,$ and $9$. This indicates the efficiency of our model in separating domain-invariant content signals. By using domain variance separation, the cross-subject classification performance is significantly improved.

\begin{figure*}[t]
    \centering\includegraphics[width=\textwidth]{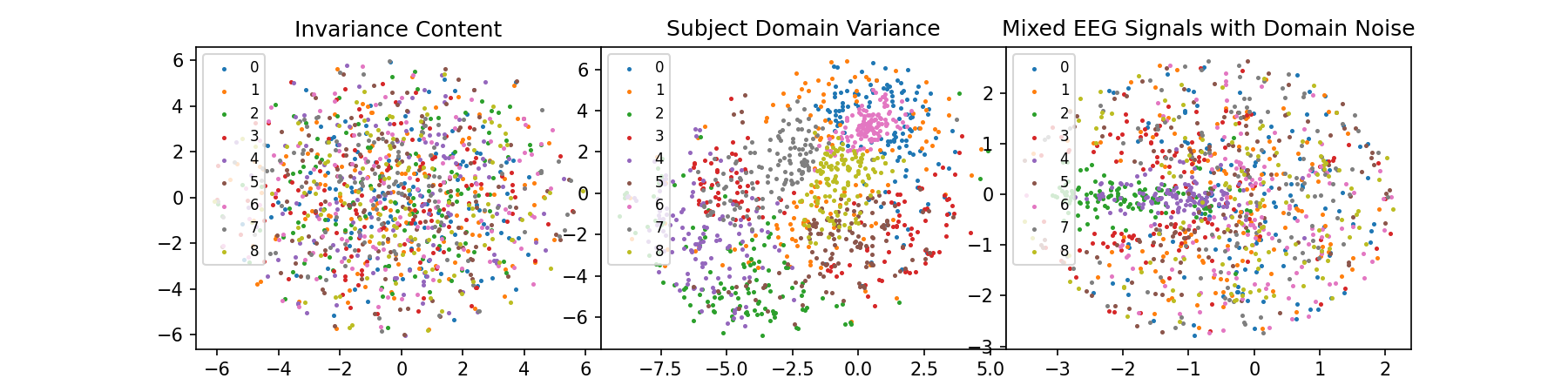}
    \caption{\label{fig:tsne-graph}T-SNE visualization of invariance content $\x^c_0$, subject domain variance $\x^s_0$, and mixed signal $\x_0$. Different color denotes different human subjects, where the invariance distribution is irrelevant to different subjects while the domain variance is clearly separable according to subject wise.  }
\end{figure*}

\subsection{Visualization of Domain Variance Distribution} 
We also visualize the domain variance distribution of each subject. 
We sample domain variance by giving different subject labels to each subject.
Then the domain variance is visualized by introducing the t-Distributed Stochastic Neighbor Embedding (t-SNE)~\cite{belkina2019automated} algorithm.
We compare the visualization of the original EEG signal $\x_0$, the separated domain variance $\x^s_0$, and the invariance content $\x^c_0$ in Figure~\ref{fig:tsne-graph}. 

As the raw signals of all channels are too large $bs \times channels \times 224 \times 75$ for t-SNE to perform significant clustering, we actually use an EEGNet pre-trained on the 4-class MI classification task, and take the mediate feature map to perform down-sampling on the raw signal. 
As the 4-class MI classification task is irrelevant to the subject information, so this downsampling does not introduce unfair bias.
It could be observed that the invariance distribution of different subjects is distributed evenly in the whole space, which illustrates the efficiency of our methods to separate ``invariance'' features from the original signal. 
For domain variance, although the degree of clustering between different classes is differentiated, the experimental results still show that the generated domain variance is strongly correlated with subjects.
These results support the efficiency of the proposed DS-DDPM method.

\section{Dicussion}
\label{subsec:abl}

\subsection{Generative Temporal Difference}

As described in the methodology, we create the network inputs by stacking segments with overlapping regions in both temporal and spatial dimensions. Consequently, the results generated by the EEG-UNet model also contain overlaps. However, for the denoising process, it is essential to have deterministic outcomes in the time domain. To tackle this issue, we incorporate a temporal difference constraint to enhance the generative consistency across the time axis. This section presents an ablation study that compares the performance of the model with and without the temporal difference constraint.

\begin{figure}[hbpt]
\vspace{-5pt}
    \centering
    \includegraphics[width=0.47\textwidth]{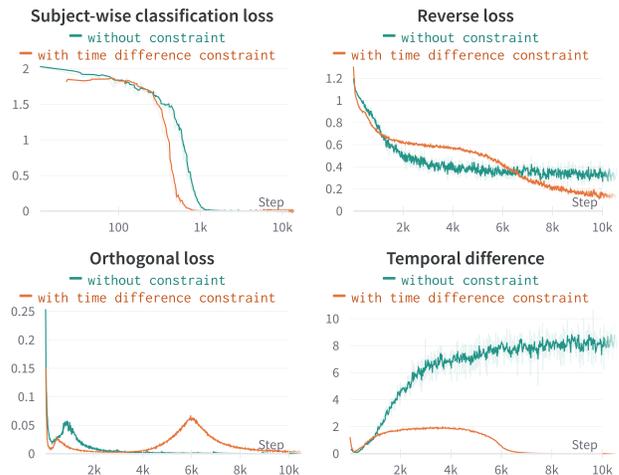}
    \caption{\label{fig:ablationlosscurve} The comparison between training metrics with or without temporal difference constraint. }
\end{figure}

We discuss the impact of incorporating temporal differences on four key metrics to assess the model's performance and report the results in Figure~\ref{fig:ablationlosscurve}. 
1) Reverse loss ($\mathcal{L}_r$), as defined in Section~\ref{subsec:domianspecific} of the main paper, measures how accurately the generative process can reconstruct the real EEG signals following the diffusion process.
2) Subject classification loss ($\mathcal{L}_{arc}$) in arc-space~\cite{deng2019arcface}, which reflects the degree of separability among individual human subjects in the extracted noise.
3) Orthogonal loss ($\mathcal{L}_{o}$) assesses the level of separability between the clean content space and noise space.
4) Temporal difference, measured using mean square error, evaluates the effectiveness of the temporal constraint in ensuring consistency across the time axis.

Figure~\ref{fig:ablationlosscurve} demonstrates that incorporating the temporal constraint effectively enables the model to learn and minimize temporal differences. In contrast, without the temporal constraint, the model generates substantial temporal differences, which are not suitable for practical applications.

Additionally, it can be observed that while the convergence speed of reverse loss is slower with the temporal constraint, it eventually outperforms the model without the constraint after 8,000 training steps, further validating the efficacy of our approach. 
Similar observations also stand for both classification loss and orthogonal loss.
Notably, implementing the temporal constraint is particularly beneficial for the classification model, as real data exhibits minimal temporal differences due to its unique and deterministic nature.

\subsection{Single Channel Denoising}

One significant advantage of DS-DDPM is its ability to perform denoising without the need for reference channels. This feature makes DS-DDPM suitable for single-channel denoising tasks. To showcase this capability, we conduct extensive experiments comparing DS-DDPM with previous methods in the context of single-channel denoising.
\begin{table}[hbpt]
\caption{Comparing DS-DDPM with traditional denoising methods \label{tb:cca_compare}}
\resizebox{0.479\textwidth}{!}{
\setlength{\tabcolsep}{1mm}{}
\begin{tabular}{ccccccccc}
\toprule
\rowcolor[HTML]{FFFFFF} 
{\color[HTML]{000000} \textbf{Sub.}} & {\color[HTML]{000000} \textbf{SCCA}} & {\color[HTML]{000000} \textbf{ECCA}} & {\color[HTML]{000000} \textbf{MsetCCA}} & {\color[HTML]{000000} \textbf{TRCA}} & {\color[HTML]{000000} \textbf{ETRCA}} & {\color[HTML]{000000} \textbf{SSCOR}} & {\color[HTML]{000000} \textbf{DS-DDPM}} & \multicolumn{1}{c}{\cellcolor[HTML]{EFEFEF}{\color[HTML]{000000} \textbf{Avg.}}} \\ \hline
\rowcolor[HTML]{FFFFFF} 
{\color[HTML]{000000} s1} & {\color[HTML]{000000} 0.50} & {\color[HTML]{000000} 0.67} & {\color[HTML]{000000} 0.50} & {\color[HTML]{000000} 0.00} & {\color[HTML]{000000} 0.00} & {\color[HTML]{000000} 0.17} & {\color[HTML]{000000} 0.29} & {\color[HTML]{000000} 0.25} \\
\rowcolor[HTML]{FFFFFF} 
{\color[HTML]{000000} s2} & {\color[HTML]{000000} 0.67} & {\color[HTML]{000000} 0.83} & {\color[HTML]{000000} 0.33} & {\color[HTML]{000000} 0.67} & {\color[HTML]{000000} 0.83} & {\color[HTML]{000000} 0.67} & {\color[HTML]{000000} 0.83} & {\color[HTML]{000000} 0.71} \\
\rowcolor[HTML]{FFFFFF} 
{\color[HTML]{000000} s3} & {\color[HTML]{000000} 1.00} & {\color[HTML]{000000} 0.83} & {\color[HTML]{000000} 1.00} & {\color[HTML]{000000} 0.83} & {\color[HTML]{000000} 0.83} & {\color[HTML]{000000} 0.83} & {\color[HTML]{000000} 0.89} & {\color[HTML]{000000} 0.88} \\
\rowcolor[HTML]{FFFFFF} 
{\color[HTML]{000000} s4} & {\color[HTML]{000000} 0.83} & {\color[HTML]{000000} 1.00} & {\color[HTML]{000000} 0.83} & {\color[HTML]{000000} 0.67} & {\color[HTML]{000000} 0.83} & {\color[HTML]{000000} 0.67} & {\color[HTML]{000000} 0.68} & {\color[HTML]{000000} 0.80} \\
\rowcolor[HTML]{FFFFFF} 
{\color[HTML]{000000} s5} & {\color[HTML]{000000} 0.83} & {\color[HTML]{000000} 0.83} & {\color[HTML]{000000} 0.83} & {\color[HTML]{000000} 0.67} & {\color[HTML]{000000} 0.67} & {\color[HTML]{000000} 0.67} & {\color[HTML]{000000} 0.67} & {\color[HTML]{000000} 0.74} \\
\rowcolor[HTML]{FFFFFF} 
{\color[HTML]{000000} s6} & {\color[HTML]{000000} 0.17} & {\color[HTML]{000000} 0.50} & {\color[HTML]{000000} 0.33} & {\color[HTML]{000000} 0.33} & {\color[HTML]{000000} 0.00} & {\color[HTML]{000000} 0.67} & {\color[HTML]{000000} 0.29} & {\color[HTML]{000000} 0.27} \\
\rowcolor[HTML]{FFFFFF} 
{\color[HTML]{000000} s7} & {\color[HTML]{000000} 1.00} & {\color[HTML]{000000} 1.00} & {\color[HTML]{000000} 1.00} & {\color[HTML]{000000} 1.00} & {\color[HTML]{000000} 0.83} & {\color[HTML]{000000} 0.83} & {\color[HTML]{000000} 0.68} & {\color[HTML]{000000} 0.87} \\
\rowcolor[HTML]{FFFFFF} 
{\color[HTML]{000000} s8} & {\color[HTML]{000000} 0.83} & {\color[HTML]{000000} 0.83} & {\color[HTML]{000000} 0.83} & {\color[HTML]{000000} 0.83} & {\color[HTML]{000000} 0.83} & {\color[HTML]{000000} 0.83} & {\color[HTML]{000000} 0.83} & {\color[HTML]{000000} 0.82} \\
\rowcolor[HTML]{FFFFFF} 
{\color[HTML]{000000} s9} & {\color[HTML]{000000} 0.33} & {\color[HTML]{000000} 0.50} & {\color[HTML]{000000} 0.33} & {\color[HTML]{000000} 0.17} & {\color[HTML]{000000} 0.17} & {\color[HTML]{000000} 0.17} & {\color[HTML]{000000} 0.50} & {\color[HTML]{000000} 0.36} \\
\rowcolor[HTML]{FFFFFF} 
{\color[HTML]{000000} s10} & {\color[HTML]{000000} 0.17} & {\color[HTML]{000000} 0.17} & {\color[HTML]{000000} 0.00} & {\color[HTML]{000000} 0.00} & {\color[HTML]{000000} 0.00} & {\color[HTML]{000000} 0.00} & {\color[HTML]{000000} 0.00} & {\color[HTML]{000000} 0.04} \\
\rowcolor[HTML]{FFFFFF} 
{\color[HTML]{000000} s11} & {\color[HTML]{000000} 0.50} & {\color[HTML]{000000} 0.83} & {\color[HTML]{000000} 0.67} & {\color[HTML]{000000} 0.83} & {\color[HTML]{000000} 1.00} & {\color[HTML]{000000} 0.67} & {\color[HTML]{000000} 1.00} & {\color[HTML]{000000} 0.82} \\
\rowcolor[HTML]{FFFFFF} 
{\color[HTML]{000000} s12} & {\color[HTML]{000000} 0.83} & {\color[HTML]{000000} 0.83} & {\color[HTML]{000000} 0.83} & {\color[HTML]{000000} 0.67} & {\color[HTML]{000000} 0.67} & {\color[HTML]{000000} 0.67} & {\color[HTML]{000000} 0.68} & {\color[HTML]{000000} {0.73}} \\
\rowcolor[HTML]{FFFFFF} 
{\color[HTML]{000000} s13} & {\color[HTML]{000000} 0.67} & {\color[HTML]{000000} 0.83} & {\color[HTML]{000000} 0.33} & {\color[HTML]{000000} 0.67} & {\color[HTML]{000000} 0.67} & {\color[HTML]{000000} 0.50} & {\color[HTML]{000000} 0.83} & {\color[HTML]{000000} 0.64} \\
\rowcolor[HTML]{FFFFFF} 
{\color[HTML]{000000} s14} & {\color[HTML]{000000} 0.17} & {\color[HTML]{000000} 0.33} & {\color[HTML]{000000} 0.33} & {\color[HTML]{000000} 0.17} & {\color[HTML]{000000} 0.33} & {\color[HTML]{000000} 0.00} & {\color[HTML]{000000} 0.33} & {\color[HTML]{000000} 0.26} \\
\rowcolor[HTML]{FFFFFF} 
{\color[HTML]{000000} s15} & {\color[HTML]{000000} 1.00} & {\color[HTML]{000000} 1.00} & {\color[HTML]{000000} 1.00} & {\color[HTML]{000000} 1.00} & {\color[HTML]{000000} 0.83} & {\color[HTML]{000000} 0.83} & {\color[HTML]{000000} 0.83} & {\color[HTML]{000000} 0.88} \\ \hline
\rowcolor[HTML]{EFEFEF} 
{\color[HTML]{000000} Avg.} & {\color[HTML]{000000} \underline{0.63}} & {\color[HTML]{000000} \textbf{0.73}} & {\color[HTML]{000000} 0.61} & {\color[HTML]{000000} 0.57} & {\color[HTML]{000000} 0.57} & {\color[HTML]{000000} 0.54} & {\color[HTML]{000000} \underline{0.63}} & \multicolumn{1}{c}{\cellcolor[HTML]{EFEFEF}{\color[HTML]{000000} }} \\ \bottomrule
\end{tabular}
}
\end{table}

We utilize a dataset~\footnote{The dataset is not publicly available due to the ongoing review process. However, the data can be provided for further verification during the reviewing process.}(BRI: Brain Robotic Interaction) consisting of single-channel EEG signals collected from human subjects controlling robot dogs using eight specific commands. This results in an eight-class classification task.
Data were collected from 15 participants, each providing 45 minutes of EEG signals.
The signals are first processed using different denoising methods and subsequently classified through a standard SSEVP classification process.
The comparison methods include widely used techniques such as SCCA~\cite{hardoon2011sparse}, ECCA~\cite{du2019multi}, MsetCCA~\cite{du2019multi}, TRCA~\cite{tanaka2013task}, ETRCA~\cite{nakanishi2017enhancing}, and SSCOR~\cite{couch1998sample}.
The results are presented in Table~\ref{tb:cca_compare}. It should be noted that for CCA-based methods, we use an additional reference channel as denoising references.
However, for DS-DDPM, denoising is performed solely using single-channel data.
DS-DDPM achieves the second-highest average accuracy across the 15 subjects without utilizing any reference signal.
This observation highlights the effectiveness of DS-DDPM in single-channel denoising tasks.

\subsection{Limitations}

The application of DS-DDPM presents a novel approach to explicitly address subject domain variance in EEG signals. However, it still faces several limitations, which we discuss in this section.

\subsubsection{Computational Complexity}

The DDPM method involves iterative processes and can be computationally expensive, particularly when dealing with high-dimensional EEG signals. This may result in longer training and inference times, as well as increased memory consumption, which might not be suitable for real-time or on-chip deployment in wearable devices. However, we note that this problem can be partially mitigated by introducing improved diffusion processes, such as DDIM and consistency models.

\subsubsection{Model Generalization}

The performance of the DDPM-based method largely depends on the quality of the training data. The DS-DDPM model alleviates the generalization problem by introducing human-subject-wise classifier guidance. However, this mechanism relies on human subjects' labels, which somewhat resembles semi-blind source separation rather than traditional blind source separation techniques used for denoising (e.g., CCA). This means that calibration data must be acquired before the model can be applied for denoising.

Currently, available datasets only contain EEG signals from tens of subjects. Drawing from experience in the computer vision and natural language processing domains, this issue could be further alleviated if the number of human subjects in EEG signal datasets was to increase on a larger scale.

\section{Conclusion}

In conclusion, this paper presents DS-DDPM, a novel conditional diffusion-denoising probabilistic model designed for domain separation in brain dynamics. By modifying the standard denoising process into two streams, we generate content signals and domain variance tailored to individual subjects. To create a feature space with desirable properties, we introduce three subtle constraints: 1) the combination of the separated streams can reconstruct the signal at each diffusion step; 2) the content feature and domain variance are orthogonal in the feature space; and 3) the domain variances are further divided according to the subjects.
Additionally, we propose UNet-EEG, a customized generative model structure specifically designed for handling long sequential samples such as EEG signals. 

The proposed DS-DDPM is not only capable of generating domain variance distributions for EEG signals from pure noise but also effectively removes human artifacts by using real EEG signals as intermediate diffusion states. 
Experimental results demonstrate the efficiency of DS-DDPM in both generating domain variance explicitly for specific subjects and enhancing cross-subject classification performance.
Overall, DS-DDPM represents a new approach to the field of brain dynamics denoising and domain separation. 
Future work may focus on further refining the model to improve its applicability across various EEG-based tasks and exploring the potential of DS-DDPM in other neuroimaging modalities.

\bibliography{ref}
\bibliographystyle{IEEEtran}
 
%

\vspace{11pt}

\newpage
\vspace{-33pt}
\begin{IEEEbiography}[{\includegraphics[width=1in,height=1.25in,clip,keepaspectratio]{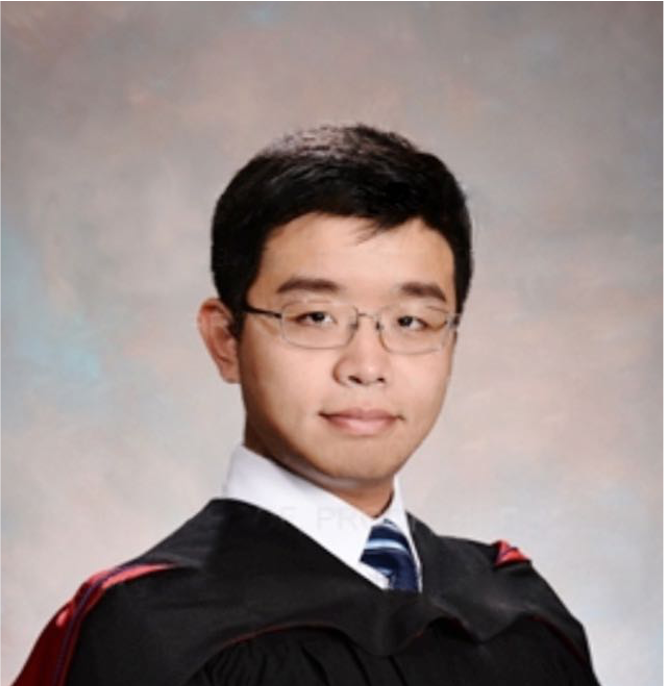}}]{Yiqun Duan} received the B.Eng. degree in electrical engineering from Hohai University, Nanjing, China, and M.A.Sc. degree in electrical engineering from the University of British Columbia, Vancouver, Canada. He is currently pursuing a Ph.D. degree in CIBCI lab, Australian Artificial Intelligence Institute, School of Computer Science, University of Technology Sydney, Australia. His current research topic is vision $\&$ language scenery understanding for embodied robotic systems and brain-computer interfaces. He is currently serving as a reviewer for top conferences in both machine learning and robotic systems. His research interests include natural language processing, reinforcement learning, and multi-model vision $\&$ language understanding.

\end{IEEEbiography}
\vspace{-33pt}

\begin{IEEEbiography}[{\includegraphics[width=1in,height=1.25in,clip,keepaspectratio]{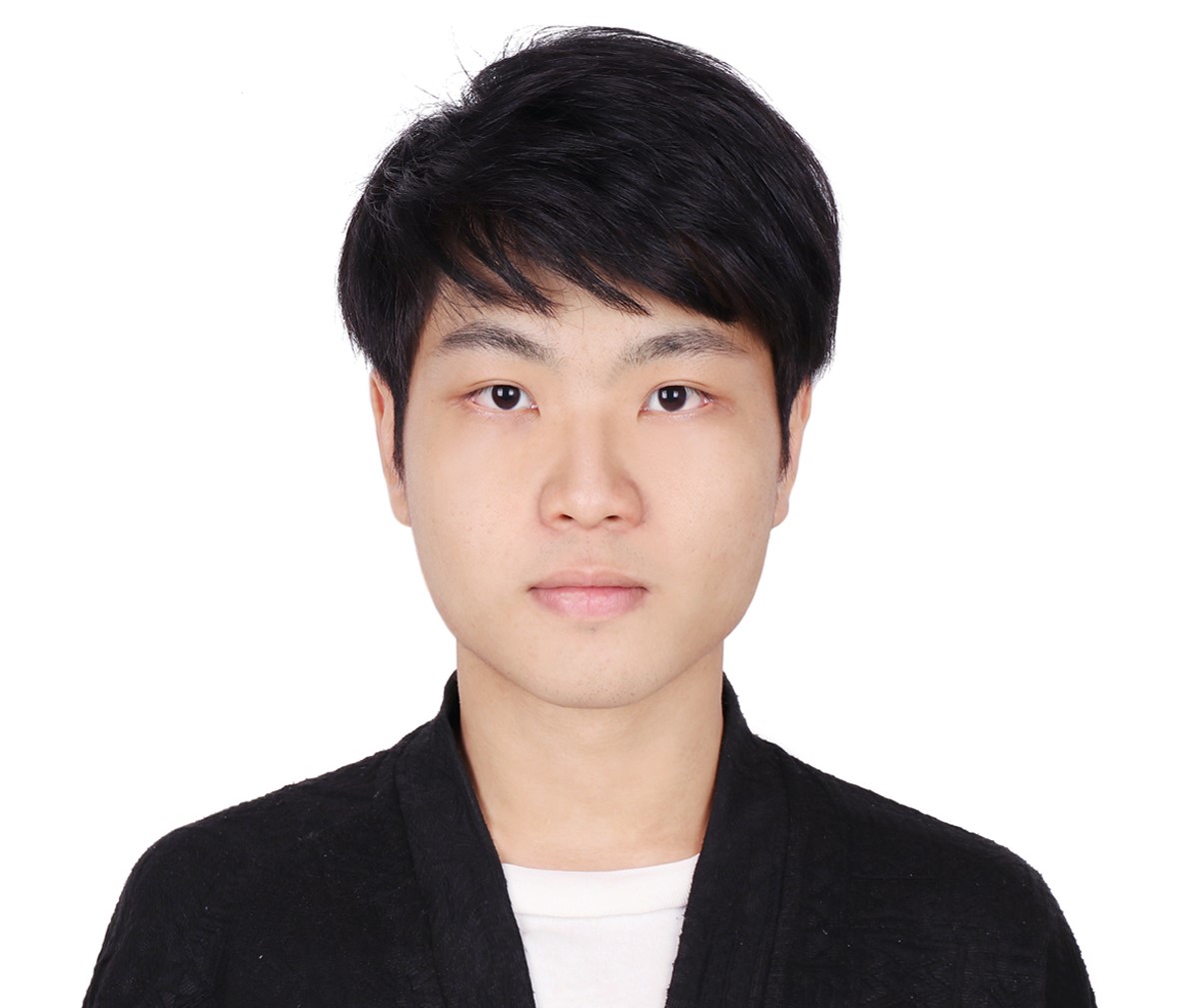}}]{Jinzhao Zhou}  received the B.Eng. degree from the School of Computer Science $\&$ Technology and the B.M.S. degree from the School of Electronic Commerce, both from the South China University of Technology (SCUT), Guangzhou, China, in 2018. He then received the M.S. degree from the School of Computer Science and Engineering from SCUT in 2021. He is currently pursuing a Ph.D. degree in CIBCI lab, Australian Artificial Intelligence Institute, School of Computer Science, University of Technology Sydney, Australia. His current research interests include Brain-Computer Interface, multi-modal understanding, and reinforcement learning.

\end{IEEEbiography}

\vspace{-33pt}
\begin{IEEEbiography}[{\includegraphics[width=1in,height=1.25in,clip,keepaspectratio]{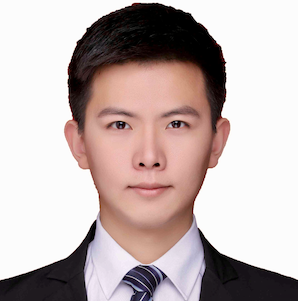}}]{Zhen Wang} received the B.Sc. degree from the Hunan University, Changsha, China, in 2016, and the M.Sc. degree from Tsinghua University, Beijing, China, in 2019. He is currently pursuing a Ph.D. degree with the School of Computer Science, The University of Sydney, Darlington, NSW, Australia. His current research interest includes streaming label learning, continual learning, and quantitative finance. He serves as a reviewer for IJCV, IEEE Trans. on Cybernetics, IEEE Intelligent Systems, ACM TKDD, NeurIPS, ICML, ICLR, CVPR, ICCV, AAAI, etc. 

\end{IEEEbiography}
\vspace{-33pt}

\begin{IEEEbiography}[{\includegraphics[width=1in,height=1.25in,clip,keepaspectratio]{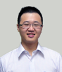}}]{Dr. Yu-Cheng Chang} received the B.S. degree in vehicle engineering from the National Taipei University of Technology, Taipei, Taiwan, in 2008, the M.S. degree with a specialization in system and control from the Department of Electrical Engineering, National Chung-Hsing University, Taiwan, in 2010 and the Ph.D. degree in software engineering from University of Technology Sydney (UTS), Australia in 2021. He currently is a Postdoctoral Researcher with the CIBCI lab, UTS. His current research interests include fuzzy systems, human performance modeling, and novel human-agent interaction.

\end{IEEEbiography}
\vspace{-33pt}
\begin{IEEEbiography}[{\includegraphics[width=1in,height=1.25in,clip,keepaspectratio]{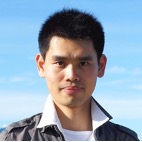}}]{Dr. Yu-Kai Wang} (M’13) received the M.S. degree in biomedical engineering and the Ph.D. degree in computer science in 2009 and 2015, respectively, both from the National Chiao Tung University, Taiwan. He is currently a Senior Lecturer at the Faculty of Engineering and Information Technology at the University of Technology Sydney, Australia. He is the author of 35 published original articles in international journals and more than 40 contributions to international conferences. His current research interests include computational neuroscience, human performance modeling, Brain-Computer Interface, and novel human-agent interaction.

\end{IEEEbiography}

\begin{IEEEbiography}[{\includegraphics[width=1in,height=1.25in,clip,keepaspectratio]{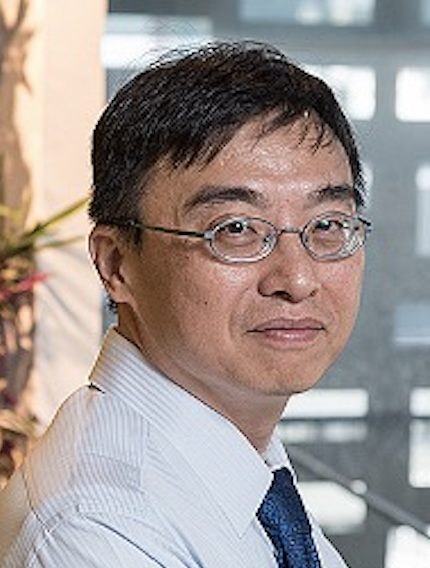}}]{Dr. Chin-Teng Lin} received a Bachelor’s of Science from National Chiao-Tung University (NCTU), Taiwan in 1986, and holds Master’s and PhD degrees in Electrical Engineering from Purdue University, USA, received in 1989 and 1992, respectively. He is currently a distinguished professor and Co-Director of the Australian Artificial Intelligence Institute within the Faculty of Engineering and Information Technology at the University of Technology Sydney, Australia. He is also an Honorary Chair Professor of Electrical and Computer Engineering at NCTU. For his contributions to biologically inspired information systems, Prof Lin was awarded Fellowship with the IEEE in 2005, and with the International Fuzzy Systems Association (IFSA) in 2012. He received the IEEE Fuzzy Systems Pioneer Award in 2017. He has held notable positions as editor-in-chief of IEEE Transactions on Fuzzy Systems from 2011 to 2016; seats on Board of Governors for the IEEE Circuits and Systems (CAS) Society (2005-2008), IEEE Systems, Man, Cybernetics (SMC) Society (2003-2005), IEEE Computational Intelligence Society (2008-2010); Chair of the IEEE Taipei Section (2009-2010); Distinguished Lecturer with the IEEE CAS Society (2003-2005) and the CIS Society (2015-2017); Chair of the IEEE CIS Distinguished Lecturer Program Committee (2018-2019); Deputy Editor-in-Chief of IEEE Transactions on Circuits and Systems-II (2006-2008); Program Chair of the IEEE International Conference on Systems, Man, and Cybernetics (2005); and General Chair of the 2011 IEEE International Conference on Fuzzy Systems. Prof Lin is the co-author of Neural Fuzzy Systems (Prentice-Hall) and the author of Neural Fuzzy Control Systems with Structure and Parameter Learning (World Scientific). He has published more than 390 journal papers including over 180 IEEE journal papers in the areas of neural networks, fuzzy systems, brain-computer interface, multimedia information processing, cognitive neuro-engineering, and human-machine teaming, that have been cited more than 28,000 times. Currently, his h-index is 77, and his i10-index is 349.

\end{IEEEbiography}

\vfill

\end{document}